\documentclass[twocolumn]{article}
\usepackage{amsmath,graphicx,amsfonts,amssymb,latexsym}
\usepackage{color}
\usepackage{subfigure}
\usepackage{fullpage}
\usepackage{subfigure}
\usepackage{newclude}
\usepackage{amsthm}
\usepackage{xkeyval}

\title{Automatic sensor-based  detection and classification of climbing activities}
\author{J\'er\'emie~Boulanger\thanks{J. Boulanger is working with the University of Rouen, France. email: {\tt jeremie.boulanger1@univ-rouen.fr} and is funded by the ANR Dynamov.}, Ludovic~Seifert\thanks{L. Seifert is working with the University of Rouen, France. email: {\tt ludovic.seifert@univ-rouen.fr}}, Romain~H\'erault\thanks{R. H\'erault is working with the INSA of Rouen, France. email: {\tt romain.herault@insa-rouen.fr}.}, Jean-Francois~Coeurjolly\thanks{JF. Coeurjolly is working with the Laboratoire Jean Kuntzmann of Grenoble, France. email: {\tt jean-francois.coeurjolly@upmf-grenoble.fr}}
}

\begin{document}
\maketitle

\begin{abstract}
This article presents a method to automatically detect and classify climbing activities using inertial measurement units (IMUs) attached to the wrists, feet and pelvis of the climber. The IMUs record limb acceleration and angular velocity. Detection requires a learning phase with manual annotation to construct the statistical models used in the cusum algorithm. Full-body activity is then classified based on the detection of each IMU. 
\end{abstract}



\section{Introduction}

The sport of rock climbing requires the use of both upper and lower limbs, as climbers reach and grasp holds and use their feet to climb the rock surface. Yet rock climbing encompasses more than continuous upward body movement, because more or less static positions are also crucial for exploring and grasping surface holds \cite{Fuss2008}\cite{Pijpers2006}\cite{Sibella2007}, posture regulation \cite{Bourdin1998}\cite{Bourdin1999}\cite{Testa1999}\cite{Testa2003}, arm release and resting \cite{Sanchez2012}, and finding routes \cite{Cordier1993}\cite{Cordier1994}. 

The time spent in exploration and posture regulation as opposed to ascending, or more broadly the time spent immobile as opposed to in motion, can be analysed by quantifying the durations when the pelvis is and is not in motion \cite{Billat1995}\cite{Sanchez2012}\cite{Seifert2013}\cite{Seifert2014}. In \cite{Billat1995}, it was noted that advanced climbers spent $63\%$ of a climb immobile and $37\%$ in actual ascending.

As previously noted, periods of immobility may not represent only inappropriate stops, but rather could reflect active resting \cite{Sanchez2012}. For example, in \cite{Fryer2012}, expert climbers were shown to spend a greater proportion of their climbing time in static states and more of the static time actively resting (i.e. limb shaking), compared with climbers of intermediate skill.

However, a point of difference with \cite{Seifert2013}\cite{Seifert2014} is that in \cite{Fryer2012} the  variability in limb behaviour was associated with exploration or a more functional use of climbing wall properties. This suggests that during stops climbers may exhibit behaviours that are dedicated to more than managing fatigue.

Ascending requires skills in route finding, which reveals the ability of climbers to interpret the ever-changing structure of the climbing wall design \cite{Cordier1993}\cite{Cordier1994}. Route finding is a critical climbing skill that can be identified by differentiating exploratory movements and performatory movements \cite{Pijpers2006}. In \cite{Pijpers2006}, a distinction between exploratory and performatory movements was made according to whether a potential hold on a climbing wall was touched, irregardless of whether it was used as a support. For example, the authors of \cite{Sibella2007} reported that skilled climbers tended to touch fewer than three surface holds before grasping the functional one.

Clearly, an excessive duration spent immobile for route finding, hold exploration or posture regulation is likely to compromise climbing fluency and lead to the onset of fatigue. The aim of this article is to propose a method to automatically detect and quantify some of the major climbing activities: immobility, postural regulation, hold exploration, hold change and traction. As stated in \cite{Seifert2013}\cite{Seifert2014}, these activities can only be defined by taking into account the activities of both the limbs and the pelvis.

\section{Data and protocol}
\label{sec_data}

\subsection{State of the art}

Previous studies like those of \cite{Billat1995} and \cite{White2010} focused on rock climbing and analysed the climber’s behaviour. This was accomplished by making video recordings of the climb and having experts manually perform the analysis. This method has several drawbacks, as the possibilities for analysis are limited, the results are of relatively low accuracy, and the process itself is long and tedious. Moreover, a full view of the climbing wall might not be available for video recording in outdoor studies without the use of drones or similar devices.

Automatic measures were reported in  \cite{Quaine1997} and \cite{Fuss2008}, with force sensors placed inside the holds on the climbing walls. Despite the high cost of the experimental devices for long routes, this method is not usable outdoors and requires a long set-up for indoor walls that are not yet equipped. 

Given the disadvantages of these methods, wireless sensors placed on the climber might be a good solution, offering easy measurement set-up and quick adaptation to the route environment (indoors or outdoors). In \cite{Pansiot2008}, for example, the climber carried a single miniature accelerometer that was used to evaluate different performance coefficients. However, this article did not present an analysis of behaviour or a procedure for distinguishing activities and their distribution along the climb. It also did not assess the activities of the different limbs, which is needed to determine the climber state on the climbing wall.
This article presents a method using multiple inertial measurement units (IMUs) placed on several body sites, with each IMU containing an accelerometer, a gyroscope and a magnetometer to detect limb and pelvic activities based on their acceleration or angular velocity. This step, presented in Section \ref{sec_cusum}, requires learning statistical modelling with a labelled set of climbs, which is accomplished by manual annotation on video recordings of climbing experts. However, these videos are no longer needed once the learning protocol is completed, and Section \ref{sec_full_body} presents a procedure for determining full-body activity by combining the independent detections of limb activity.


\subsection{Protocol}

Two male climbers of ability 6a on the French Rating Scale of Difficulty (F-RSD) \cite{Delignieres1993}, which corresponds to an intermediate performance level  \cite{Draper2011}, undertook an easy, top-roped route (grade of 5c on F-RSD) composed of 20 hand holds for a 10m height. The route was identifiable by colour and set on an artificial indoor climbing wall by two certified route setters who ensured that it matched an intermediate level of climbing performance. The participants were instructed to self-pace their ascent and to climb fluently and without falling. The ascents were preceded by 3 minutes of route preview, as pre-ascent visual inspection is a key parameter of climbing performance  \cite{Sanchez2012}. Procedures were explained to the climbers, who then gave written informed consent to participate1.\footnote{The protocol was approved by the local University ethics committee and followed the declaration of Helsinki.}.

Each climber was considered to be in one the following states at any given time: immobility, postural regulation, hold exploration, hold change or traction. As stated previously, a single detection of limb activity is required to describe the full-body state.

Accelerations and angular velocities were collected from the four limbs and pelvis using IMUs located on the right and left wrists, right and left feet, and pelvis. The IMUs combined a triaxial accelerometer ($\pm8G$), a triaxial gyroscope ($1600^\circ/s$) and a triaxial magnetometer (MotionPod, Movea\textcopyright, Grenoble, France) referenced to magnetic North, sampled at 100Hz. Wireless transmissions to a controller enabled recording with MotionDevTool software (Movea\textcopyright, Grenoble, France). A wearable device is required to measure climbing activities on a $10$m-high climbing wall.


\subsection{Recording and preprocessing}

The acceleration of each sensor was determined from the recorded signals. These data were used to synchronise the video recording of the climb with the IMU signals in the learning phase of the detection, as well as to detect  activity. Therefore, this method had to be used even on unlabelled sets of climbing.

Although the sensors directly record the acceleration in the sensor frame, the recording cannot be used directly as proper acceleration because of the gravity component. The norm of this component is well-known ($9.81m/s^2$), but it cannot be removed from the sensor frame without knowing the orientation of the sensor in the ground reference frame.

\begin{figure}
\centering
\includegraphics[width=1\linewidth]{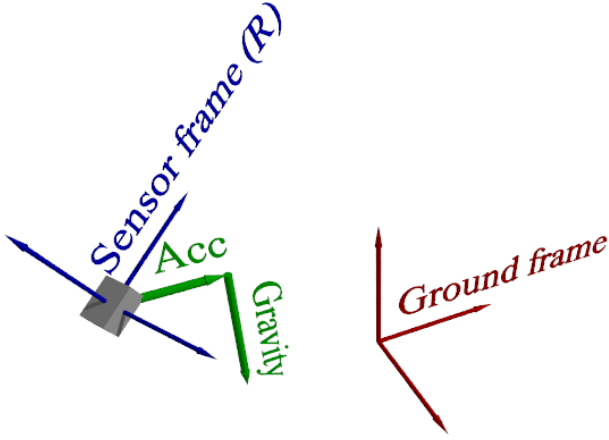}
\caption{Example of the frame difference between the ground frame (red), where the gravity is known, \textit{i.e} can be removed, and the sensor (grey box) frame (blue), where the gravity components are unknown and cannot be removed from the recorded acceleration (sum of the green vectors).
\label{fig_frame}}
\end{figure}

Let $a_s$ be the recorded acceleration in the sensor frame. By denoting $R$ the rotation matrix describing the sensor frame in the Earth reference frame (magnetic North, West and vertical up direction), the acceleration a in the Earth reference is then defined as $a=Ra_s$. Once $a$ is obtained, the gravity component can easily be removed (see Figure \ref{fig_frame}).

To determine $R$, a complementary filter based algorithm is used \cite{Madgwick2010}\cite{Madgwick2011}, based on the three sensor information sources (i.e. accelerometer, gyroscope and magnetometer). The gyroscope measures precise angular changes over very short time durations but cannot be used to track the angle changes by integration due to drift. The accelerometer provides absolute, albeit noisy, measurements of acceleration. By combining the two sensor information sources it was possible to reduce the drift of the gyroscope for sensor orientation tracking. When magnetometer information was added, it was possible to compute the sensor orientation respect to the fixed frame of Earth reference. 

\begin{figure}
\centering
\includegraphics[width=1\linewidth]{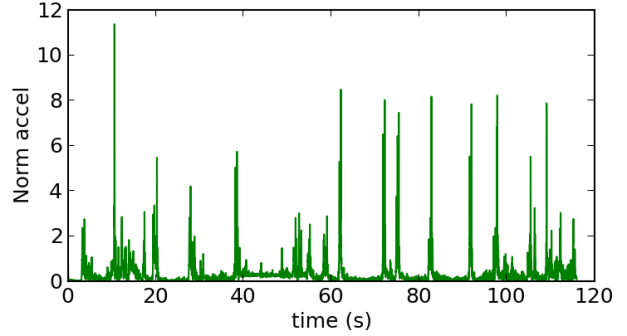}
\caption{Example of the evolution in the norm of a recorded acceleration after removing the gravity component. The sensor was attached to the climber’s right foot. The acceleration is measured in $m/s^2$.
\label{fig_accel_example}}
\end{figure}

Figure \ref{fig_accel_example} provides an example of the recorded norm of the processed acceleration signal.

Based on the processed acceleration and the angular velocity, we automatically detected the limb motion using the cusum \cite{Granjon2012} method.

\section{Activity detection}
\label{sec_cusum}

Activity was detected from each sensor independently. As this method requires a learning phase, this last is also described in detail. Once the learning phase is completed, the paragraph \ref{subsec_cusum} can directly be used.

\subsection{Cusum-based detection}
\label{subsec_cusum}

The cusum algorithm is often used to determine a change point in a time series of independent random variables based on statistical models. In this case, we assume that a limb is either immobile (state called $H_0$) or mobile (in motion) (state $H_1$) at a given time. It is further assumed that these states are exhaustive and exclusive, i.e. for each sample, the limb is in one of these states and only one. The idea is thus to estimate the state of the limb based on the recorded signals. Let $x_t$ be the considered signal. In our case, it will be the norm of either the acceleration or the angular velocity.

The main idea is to assume that $x_t$ is a random variable sampled from a distribution fixed by the state of the limb. In other words, if the limb is in state $H_0$ (respectively $H_1$), then $x_t\sim p(.|H_0)$ (respectively $x_t\sim p(.|H_1)$). If $p(.|H_0)$ and $p(.|H_1)$  are known, a likelihood ratio can be computed for a given $t$ (we directly consider $t$ as a discrete variable, due to the sampling process) to estimate the distribution from which $t$ was sampled
$$
l_t^x=\log\left(p(x_t|H_1)\right)-\log\left(p(x_t|H_0)\right).
$$
The sign of $l_t^x$ provides an estimation of the state of the limb at time $t$: $H_1$ for positive and  $H_0$ for negative.

It is assumed that the state changing periods are much longer than the sampling time and can therefore form a cumulative sum to increase the detection performance. Let $S_t^x$ be the cumulative sum of the log likelihood ratio
\begin{equation}
S_t^x=\sum_{k\leq t} l_k^x.
\label{eq_cumsum}
\end{equation}
The propensity of $S_t^x$ to be monotonous indicates the state the limb should be assumed to be in. A change in monotony implies a change in the state. An algorithm looking directly at a change in monotony would be subject to many false detections. Instead, the use of positive thresholds $\lambda_0$ and $\lambda_1$  helps to reduce false alarms.

For example, given  state $H_0$ at $t=0$, state $H_1$ is detected when
$$
S_t^x>min(S_s^x|s<t)+\lambda_1.
$$
Similarly, to detect $H_0$ while being in  state $H_1$, a change occurs when
$$
S_t^x<max(S_s^x|s<t)-\lambda_0.
$$
When a change point is detected, the process starts again by taking the detection time as the new time origin and starting the cumulative sum again from this point.

Although thresholds can be chosen to be equal, we consider a more general framework here by taking different values. These thresholds can influence the false positive detection rate and therefore should be chosen according to a given performance measure or some prior model of detection.

Thresholds as well as the distributions $p(.|H_0)$ and $p(.|H_1)$ are not known in advance in this case, which is why a learning step is required. A labelled set is used to estimate the distributions and thresholds based on manual annotation, described in the next section.

\subsection{Construction of the labelled set}

The climbs were video recorded in their entirety, and experienced climbers manually annotated three different climbs by two different climbers. They indicated for each frame of the video the state of each limb.

\begin{figure}
\centering
\includegraphics[width=0.8\linewidth]{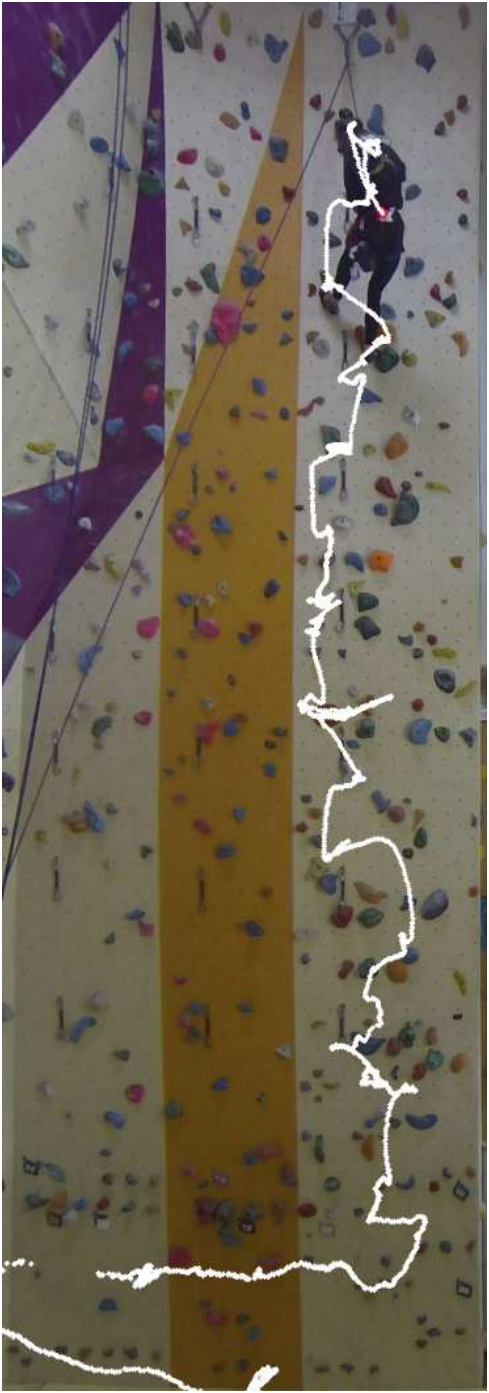}
\caption{Example of automatic tracking of the climber based on video footage. The trajectory during the climb is shown in white. The background image is the last image from the video, therefore containing the last tracked position. It is also apparent that the red light can be hidden, for example, due to substantial body rotation so that the light no longer faces the camera. On this image, the lens distortion and the parallax were corrected. 
\label{fig_tracking}}
\end{figure}

Because the camera and sensors were not synchronised, the delay between the frame-based manual annotations and the acceleration had to be estimated. The videos were recorded using a fixed camera facing the wall, with a red light attached to the climber’s pelvis. The position of the red light on the image therefore gave the position of the pelvis on the wall, up to some correction. A classical Kalman filter was used to track the red light in the frame, and an adaptive filter was required to counterbalance similar colours in the surrounding environment (in this case, the presence of televisions in the recorded picture). Once the red-light position on the image was obtained, lens distortion had to be corrected, followed by parallax correction. An example of automatic tracking after correction is presented in Figure \ref{fig_tracking}.

\begin{figure}
\centering
\includegraphics[width=0.4\textwidth]{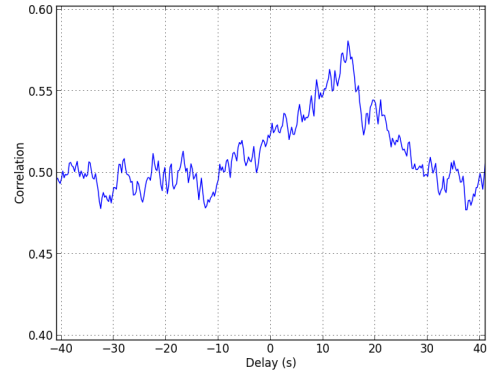}
\caption{Example of the correlation between the sensor acceleration and the video tracking-based acceleration. In this case, the delay maximising the correlation is around $14.7s$.
\label{fig_correlation}}
\end{figure}

Based on the obtained trajectory, an approximate acceleration of the pelvis was determined. Using a maximum correlation measure with the sensor-based lateral and vertical accelerations of the pelvis, the delay between each signal was estimated. Therefore, the manual annotation (based on the video) and the recorded signals (based on the sensors) could be synchronised. An example of correlation is presented in Figure \ref{fig_correlation}.

An example of the synchronised annotation and signal is presented in Figure \ref{fig_annotation}. Based on these annotations, a model was determined for each state. In this example, it appears clearly that the high values of the accelerations match the $H_1$ hypothesis (mobility). These three annotated climbs, along with the recorded IMU signals, will be used as a labelled set for the learning phase of activity detection.

\subsection{Learning protocol}

\begin{figure}
\centering
\includegraphics[width=0.4\textwidth]{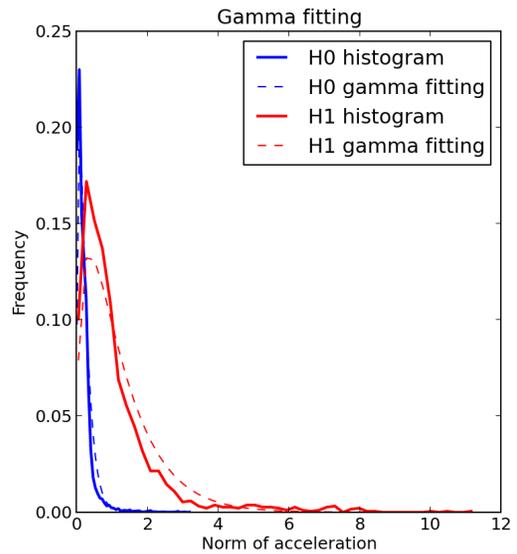}
\caption{Comparison between the histogram of the acceleration (solid line) and the fitting Gamma distribution (dotted line). For each state, only the corresponding (based on the annotations from Figure \ref{fig_annotation}) samples are considered. The fitting Gamma distribution is trying to fit the histogram and is obtained via a maximum likelihood estimation.
\label{fig_modele}}
\end{figure}

For now, we consider that $x_t\in \mathbb{R}^+$ is either the norm of the acceleration or the norm of the angular velocity. Using the manual annotations, an example of the histogram of the acceleration norm for each hypothesis $H_0$ and $H_1$ is presented in Figure  \ref{fig_modele}.

Instead of considering general distributions for $p(.|H_i)$, $i=0,1$, parametric distributions are considered. A $\chi^2$ test is performed for each sensor and each climb and the results are presented in Table \ref{table_chi2}. The values presented in Table \ref{table_chi2} are averages of the p-values for the three annotated climbs. Table \ref{table_chi2} compares the fit between the norm of the acceleration and its squared value to the Gamma distribution, the exponential distribution (which seems to fit the samples in the $H0$ hypothesis), and the $\chi^2$ distribution with three degrees of freedom (corresponding, for the squared norm, to a model where each component of the signal is sampled from a Gaussian). It indicates that the Gamma distribution-based model seems a good choice to describe the signals.

\begin{table}
\centering
\begin{tabular}{|c|c|c|c|c|}
\hline
& \multicolumn{2}{c|}{Acceleration}&\multicolumn{2}{c|}{Ang. velocity}\\ \hline
&norm&squared &norm&squared\\ \hline
Gamma $H0$& 1.0	& 0.68	&  1.0	& 0.93 \\ \hline	
Expone $H0$& 1.0	& 0.16	& 1.0	& 0.53 \\ \hline	
Gamma $H1$& 0.59	& 0	& 0.93 & 0.32 \\ \hline	
$\chi^2$ $H1$& 0	& 0.13	& 0.28	& 0.59 \\ \hline	
\end{tabular}
\vspace{1mm}
\caption{p-values of $\chi^2$ goodness-of-fit tests for the norm and squared norms of the acceleration and angular velocity signals. The Gamma and exponential models are tested for state $H_0$, while Gamma and  with 3 degrees of freedom models are tested for state $H_1$.
\label{table_chi2}}
\end{table}


To be more specific, we consider the Gamma distributions $p(.|H_i)$ for $i = 0, 1$ for both states given by
\begin{equation}
p(x|H_i)=\frac{1}{\Gamma(k_i){\theta_i}^{k_i}}x^{k_i-1}\exp\left(\frac{-x}{\theta_i}\right),
\label{eq_gamma}
\end{equation}
where $\theta_i$ and $k_i$ are positive real number and $i=0,1$ represents the state.

The coefficients $\theta_i$ and $k_i$ are determined by maximum likelihood. By considering a second order approximation of the digamma function $\Gamma'(k)/\Gamma(k)$ \cite{Minka}, they are given by
\begin{equation}
\hat{k_i}\approx\frac{3-s_i+\sqrt{(s_i-3)^2+24s_i}}{12s_i}
\label{eq_k}
\end{equation}
and
\begin{equation}
\hat{\theta_i}=\mathbb{E}_i[x]/k_i,
\label{eq_theta}
\end{equation}
with
$$
s_i=\log\left(\mathbb{E}_i[x]\right) - \left[\mathbb{E}_i[\log(x)\right].
$$
The term $\mathbb{E}_i[x]$ represents the empirical average of the values of $x$ when the annotation indicates it is in state $i$. These empirical means are determined by considering the concatenation of the signals from the labeled set.

Clearly, these parameters change with the nature of the signal $x_t$ (norm of the acceleration or norm of the angular velocity).

\begin{figure}
\centering
\includegraphics[width=1\linewidth]{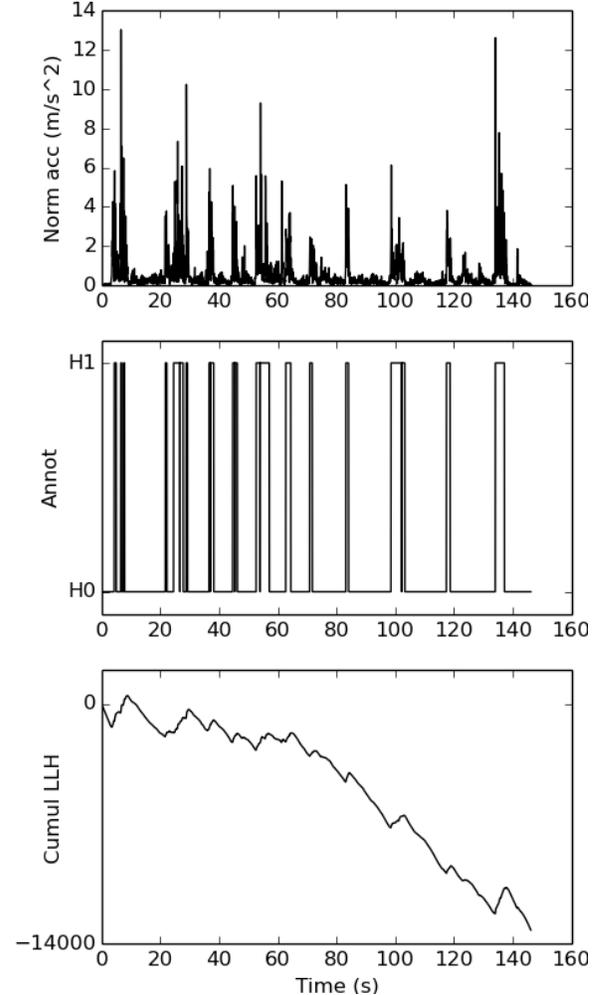}
\caption{(Top) Example of the norm of the acceleration for one sensor. (Middle) Manual annotation synchronized with the acceleration signal. (Bottom) Cumulated loglikelihood $S_t^{acc}$ defined in Equation (\ref{eq_cumsum}) of the acceleration signal using the models constructed from the annotation.
\label{fig_annotation}}
\end{figure}

Based on the signal $x_t$, using these Gamma distributions with parameters determined via Equations (\ref{eq_k}) and (\ref{eq_theta}), the cumulated log likelihood ratio $S_t^x$ can be determined. An example is presented, along with the annotations, in Figure \ref{fig_annotation}.\\

The method presented in Paragraph \ref{subsec_cusum} leads to the determination of a cumulated log likelihood ratio (see Equation (\ref{eq_cumsum})), depending on the given signal. Let $S^{acc}_t$ be based on the norm of the acceleration and $S^{ang}_t$ be based on the norm of the angular velocity. 

This is due to a lack of proper multivariate Gamma model, preventing the use of both signals directly. Instead, a basic solution is used here where the cusum algorithm is run on $S_t$, obtained as a weighted sum of  $S^{acc}_t$ and  $S^{ang}_t$:
\begin{equation}
S_t=\alpha S^{acc}_t+(1-\alpha) S^{ang}_t
\label{eq_ponderated}
\end{equation}
where $\alpha\in[0,1]$.
For example, if $\alpha=0$ (respectively $1$), then only the angular velocity (respectively acceleration) is considered for detection. A comparaison of the performances depending on $\alpha$ is presented in Section \ref{sec_perf_lim}. It should be noted that although  $S^{acc}_t$ and  $S^{ang}_t$  are cumulated log likelihood ratios, $S_t$ is no longer one as the only possible distribution associated with such a likelihood would not be unitary. Therefore, maximising (\ref{eq_ponderated}) does not correspond to maximising a likelihood. The minimisation of $S_t$ should be regarded as a minimum contrast estimation method.

\begin{figure}
\centering
\includegraphics[width=1\linewidth]{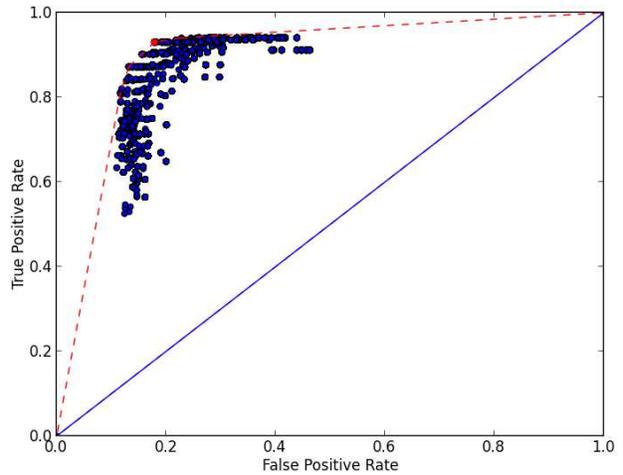}
\caption{Example of a ROC figure, representing the true positive rate with respect to the false positive rate. Each dot represents an estimation for a given set of thresholds. The best set of thresholds maximising (\ref{eq_perf}) is the one maximising the distance to the diagonal. In this example, the norm of the angular velocity was the signal used for detection and the sensor was attached to the left foot.
\label{fig_roc}}
\end{figure}

Running the cusum algorithm with different thresholds $\lambda_i$ will clearly lead to different estimates. To measure the performance of the detection, the coefficient 
\begin{equation}
\label{eq_perf}
c=\frac{TP}{P}-\frac{FP}{N}
\end{equation}
is used, where $TP$ is the number of true positives (detecting $H_1$  when it is $H_1$), $P$ (respectively $N$) the number of elements in $H_1$ (respectively $H_0$) in the manual annotation, and $FP$ the number of false positives (detecting $H_1$ when it is $H_0$). This coefficient corresponds to the performance measure used in the learning protocol as it represents, in a ROC curve, twice the distance to the diagonal, indicating a random decision (Figure \ref{fig_roc}). The thresholds are then determined as the ones maximising $c$.

\begin{figure}
\centering
\includegraphics[width=1\linewidth]{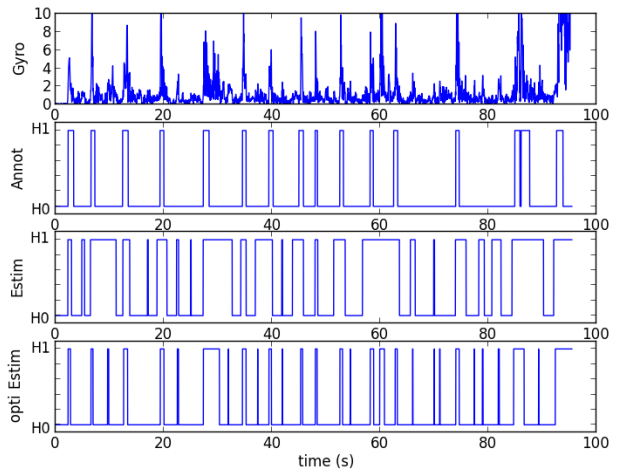}
\caption{(Top) Norm of the angular velocity. (Middle top) Manual annotation. (Middle bottom) Detection. (Bottom) Optimal estimation. In this case, the sensor was attached to the right hand. For better readability, the weighted coefficient $\alpha=0$ was chosen; therefore, only the angular velocity was used in the determination of $S_t$. It is notable that the optimal estimation still differs from the annotation. \label{fig_detect_optim}}
\end{figure}

A comparison of the recorded signal, the annotation and the (optimal) detection is presented in Figure \ref{fig_detect_optim}. This figure also illustrates the difference between the detection based on other annotations where all the parameters are learnt from a distinct labelled set and the optimal detection based on the signal using the annotation of the signal itself to perform the detection. This optimal detection is not achiveable in practice as it requires the annotation of the considered climb.

The next section presents the performance of this detection via a cross-validation method.

\subsection{Performance and limitations}
\label{sec_perf_lim}

A cross-validation method was used to evaluate the performances of the learning algorithm. The method consists of learning the different parameters ($k_i$, $\theta_i$, $\lambda_i$ for $i=0,1$), concatenating two out of the three annotated climbs, and applying these parameters on the third one. The annotation for the third climb makes it possible to determine a performance measure via the coefficient from (\ref{eq_perf}). An optimal coefficient $c$ can also be determined by directly learning all the parameters only using the third climb. This gives an idea of the best achievable performance. Then, permutations between the learning and testing climbs are made and the average coefficient $c$ is considered as the final performance of the algorithm. Table \ref{table_perf} presents the results from the cross-validation for all the sensors for $\alpha=0,1$ and the value maximising the score (and therefore, better than the extrema value). It appears that the score is roughly similar for the different values of $\alpha$. However, the optimal value of $\alpha$ (different for each sensor) will be used from now on. For further applications with non-annotated climbs (see Paragraph \ref{subsec_applic}), the learning phase is carried out on all three annotated climbs.

\begin{table}
\centering
\begin{tabular}{|c|c|c|c|c|c|c|}
\hline
& \multicolumn{2}{c|}{Accel.}&\multicolumn{2}{c|}{Ang. vel.}&\multicolumn{2}{c|}{Acc, Ang vel}\\ \hline
&score&opt&score&opt&score&opt \\ \hline
L F& 0.79	& 0.80 & 0.80 & 0.81	&  0.67	 & 0.82\\ \hline	
R F& 0.74	& 0.80 & 0.75 & 0.83	& 0.64	 & 0.84\\ \hline	
L H& 0.47	& 0.62 & 0.53 & 0.64	& 0.48  & 0.68\\ \hline	
R H& 0.42	& 0.53 & 0.43 & 0.60	& 0.37	&  0.59\\ \hline	
Pelv.& 0.31	& 0.38 & 0.34 & 0.48	& 0.11 & 0.49\\ \hline	
\end{tabular}
\vspace{1mm}
\caption{Score and optimal score of the cross-validation for $\alpha=0$ (Ang. velocity), $\alpha=1$ (Accel) or the optimal $\alpha$. The score is only slightly increased for the optimal value.
\label{table_perf}}
\end{table}

The optimal detection did not provide a very high performance measure in all cases. This may have had several causes:
\begin{itemize}
\item \underline{Missing a movement during manual annotation}: Due to the lack of visibility of the concerned limb. 
\item
\underline{Delay or different movement period}: As the annotation is manual, a delay might occur between a movement and its detection by the person annotating the video.
\item
\underline{Sensor is hit during the climb}: For example, this occurs when the climber claps his hands together, creating an acceleration peak of both wrist sensors.
\item
\underline{Defective orientation estimation}: As the acceleration signal requires the sensor orientation, a wrong orientation estimation will directly add bias to the acceleration because the gravity component will no longer be aligned with the vertical (according to the sensor).
\end{itemize}

The next section presents how these binary detections (state $H_1$ or $H_0$) from each sensor can be classified to describe a full-body state and how this is used to measure  exploration during a climb.


\section{Activity classification}
\label{sec_full_body}

\subsection{Full-body activity}
\label{subsec_fb}

Based on state detection of the four limbs and pelvis, we defined four exclusive states matching the different activities of the climber:
\begin{itemize}
\item
\underline{Immobility}: All limbs are immobile and the pelvis is immobile.
\item
\underline{Postural Regulation}: All limbs are immobile and the pelvis is moving.
\item
\underline{Hold interaction}: At least one limb is moving and the pelvis is immobile.
\item
\underline{Traction}: At least one limb is moving and the pelvis is moving.
\end{itemize}

Immobility is the state when the climber is not moving at all. He might be resting due to fatigue or looking at the route to determine its climbing path, while his limbs remain immobile.

Postural regulation is the adjustment of the climber’s centre of mass while his limbs stay on the same holds. This might consist of a body rotation to be able to catch a hold that would not be reachable with the previous body configuration.

Hold interaction is the movement of a limb while maintaining the pelvis (and therefore the global position of the climber on the wall) immobile. This is a change in the hold in use before the next traction, a change in the position and orientation of the hand/foot on an hold for better adapted use of the hold, or successive limb movement to determine which hold is most appropriate for the next traction. In the next section, we present a more detailed classification to differentiate actual use of a hold from hold exploration.

Traction is the state when the climber is moving (generally upward) using at least one limb. Although the limb might not be moving as substantially as during a hold change, this state is still easily detected and its definition seems to fit the climber’s actual traction phases on the video.

An example of the successive states for a climb is presented in the last panel of Figure \ref{fig_affordance}.

\subsection{Limb state}

\begin{figure}
\centering
\includegraphics[width=0.9\linewidth]{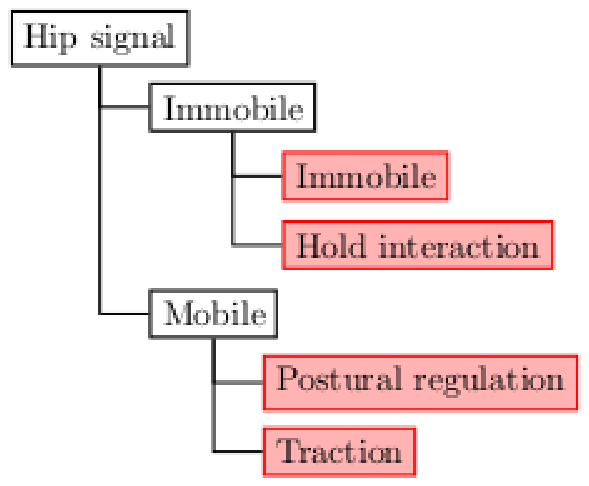}
\includegraphics[width=1\linewidth]{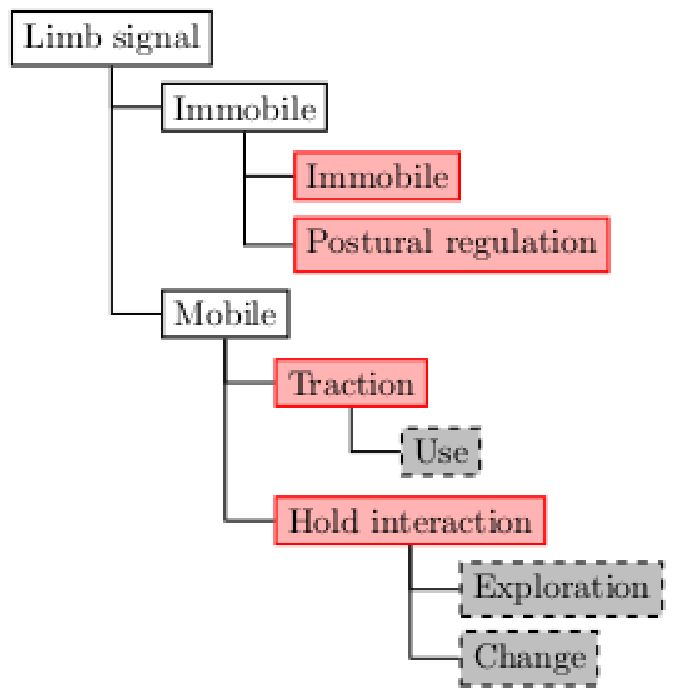}
\caption{Decision tree from the signals. White rectangles are decisions from single sensors and red rectangles are multiple sensor decisions. Each decision depends on the decisions from other senors. The last layer for the limb detection tree depends on the next traction detection.
\label{fig_tree}}
\end{figure}

\begin{figure}
\centering
\includegraphics[width=1\linewidth]{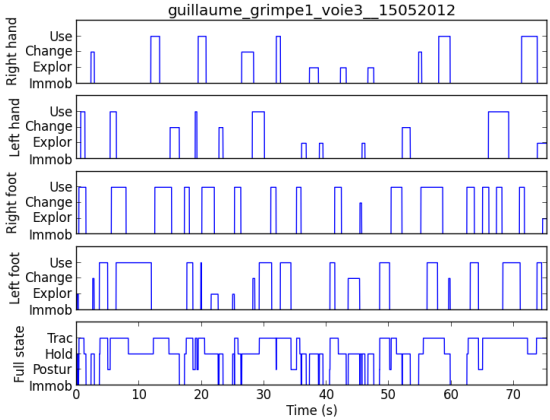}
\caption{Example of classification from a non annoted climb. From top to bottom: full state evolution for each limb: right hand, left hand, right foot, left foot and full body state evolution during the climb.  
\label{fig_affordance}}
\end{figure}

It quickly appears that the state of hold interaction covers a wide spectrum of activities. It is not directly possible to determine whether the climber is actually using a hold or moving a limb towards different holds to check whether they are reachable and, if so, how to use them (in this case, the limb can remain on the same hold but change its orientation). Consequently, new states for each limb are considered as sub-states of the Hold interaction defined in Section \ref{subsec_fb}. These sub-states are the following:


\begin{itemize}
\item
\underline{Immobility}: When a limb is detected as being immobile.
\item
\underline{Use}: When a limb is moving during traction.
\item
\underline{Change}: The last movement before traction, or the final change in hold (or change in limb orientation on the same hold) before being used.
\item
\underline{Exploration}: All movements except the last one before traction. An example is the case when the climber is trying several holds before choosing the one he will be using for traction.
\end{itemize}
A summary of the different states is presented in Figure \ref{fig_tree} and an example of the full state analysis of a climb is presented in Figure \ref{fig_affordance}.

\subsection{Example of application}
\label{subsec_applic}

This section presents a simple example of application based on the previous algorithm: measuring the ratio between exploratory and performatory movements.

\begin{figure}
\centering
\includegraphics[width=1\linewidth]{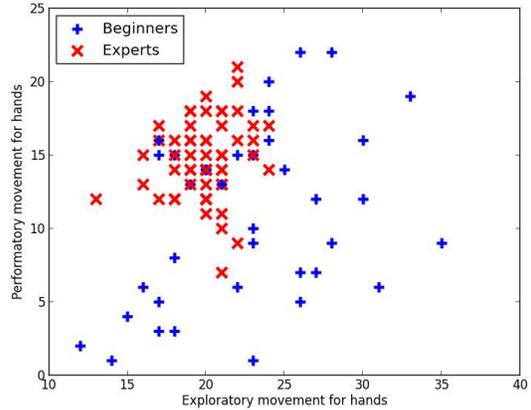}
\caption{Comparison between the number of holds touched by the hands (exploratory movement) and the number of holds used (performatory movement) during a climb. Results for the feet are quite similar. A classification clearly appears based on the ratio between these quantities. 
\label{fig_touch_hand}}
\end{figure}

As previously noted, this ratio can be used to differentiate skilled and unskilled climbers \cite{Sibella2007}, which makes automatic detection of the ratio very useful. Based on the present algorithm, {\it Exploration} and {\it Change}  encompass all the exploratory movements for each limb, whereas {\it Use} reflects only the performatory movements. The results based on a total of 94 climbs by three experts and three beginners on a 10m-high climbing route are presented in Figure \ref{fig_touch_hand}. A distinction between experts and beginners clearly emerges based on the ratio between these quantities. 

Like previously stated, it has been shown \cite{Sibella2007} that this ratio can be used to differentiate skilled and unskilled climbers. There is therefore an interest in automatically detecting this ratio. Based on the present algorithm, the number of exploratory movements will gather, for each limb, the number of \textit{Exploration} and \textit{Change} while the number of performatory movement only counts the \textit{Use} activity periods. Results based on a total of 94 climbs realized by 3 experts and 3 beginners on a 10 meters high climbing route are presented in Figure \ref{fig_touch_hand}. A distinction between experts and beginners clearly appears considering the ratio between these quantities 

It should be noted that some of the values from this figure seem rather high for a 10m-high climbing route. This is partly due to the computation of the sensor orientation, which is still somewhat inexact. If the orientation is not properly determined, the computed acceleration contains residuals from the gravity component, inducing a {\it mobile} phase detection further on and misleading the counting process. One possibility to eliminate this issue might be to only consider the angular velocity value, as the performances do not differ significantly between angular velocity and acceleration use. Another factor might be individual differences between the climbers from the labelled set used for the learning protocol and the climbers used in this application. To prevent this individual dependent measure, the learning phase can be accomplished using a dedicated climber recording and the learnt parameters only for this specific climber.


\section{Conclusion}

This article presents a method for the automatic detection and classification of climbers’ activities based on multiple IMUs. From a learning phase requiring manual annotations, a statistical model is built for the norms of acceleration and angular velocity. This model is used in a cusum algorithm to detect a binary movement state for each limb with an attached sensor. The concatenation of the states of each sensor is used to determine and classify full-body activity. A more detailed classification is then used to measure exploration during the climb. 

Determining exploratory activity during climbing is useful as it provides a measure of skill and learning performances. Future works will focus on the use of this method in a learning protocol for indoor climbing, measuring the occurrence and distribution of exploration and immobility in the participants. Another study on immobility is planned to determine which body configurations are mainly used during learning and how these configurations evolve from climb to climb during a learning protocol


\begin{thebibliography}{10}
\providecommand{\url}[1]{#1}
\csname url@samestyle\endcsname
\providecommand{\newblock}{\relax}
\providecommand{\bibinfo}[2]{#2}
\providecommand{\BIBentrySTDinterwordspacing}{\spaceskip=0pt\relax}
\providecommand{\BIBentryALTinterwordstretchfactor}{4}
\providecommand{\BIBentryALTinterwordspacing}{\spaceskip=\fontdimen2\font plus
\BIBentryALTinterwordstretchfactor\fontdimen3\font minus
  \fontdimen4\font\relax}
\providecommand{\BIBforeignlanguage}[2]{{%
\expandafter\ifx\csname l@#1\endcsname\relax
\typeout{** WARNING: IEEEtran.bst: No hyphenation pattern has been}%
\typeout{** loaded for the language `#1'. Using the pattern for}%
\typeout{** the default language instead.}%
\else
\language=\csname l@#1\endcsname
\fi
#2}}
\providecommand{\BIBdecl}{\relax}
\BIBdecl

\bibitem{Fuss2008}
F.. Fuss and G.. Niegl, ``Instrumented climbing holds and performance analysis
  in sport climbing,'' \emph{Sports Technology}, vol.~1, pp. 301--313, 2008.

\bibitem{Pijpers2006}
J.~R.. Pijpers, R.~D.. Oudejans, F.. Bakker, and P.~J.. Beek, ``The role of
  anxiety in perceiving and realizing affordances,'' \emph{Ecological
  Psychology}, vol.~18, pp. 131--161, 2006.

\bibitem{Sibella2007}
F.. Sibella, I.. Frosio, F.. Schena, and N.~A.. Borghese, ``3d analysis of the
  body center of mass in rock climbing,'' \emph{Human Movement Science},
  vol.~26, 2007.

\bibitem{Bourdin1998}
C.. Bourdin, N.. Teasdale, and V.. Nougier, ``High postural constraints affect
  the organization of reaching and grasping movements,'' \emph{Experimental
  Brain Research}, vol. 122, pp. 253--262, 1998.

\bibitem{Bourdin1999}
C.. Bourdin, N.. Teasdale, V.. Nougier, C.. Bard, and M.. Fleury, ``Postural
  constraints modify the organization of grasping movements,'' \emph{Human
  Movement Science}, vol.~18, pp. 87--102, 1999.

\bibitem{Testa1999}
M.. Testa, L.. Martin, and B.. Deb\^u, ``Effects of the type of holds and
  movement amplitude on postural control associated with a climbing task,''
  \emph{Gait and Posture}, vol.~9, pp. 57--64, 1999.

\bibitem{Testa2003}
------, ``3d analysis of posturo-kinetic coordination associated with a
  climbing task in children and teenagers,'' \emph{Neuroscience Letters}, vol.
  336, 2003.

\bibitem{Sanchez2012}
X.. Sanchez, P.. Lambert, G.. Jones, and D.~J.. Llewellyn, ``Efficacy of
  pre-ascent climbing route visual inspection in indoor sport climbing,''
  \emph{Scandinavian Journal of Medicine and Science in Sports}, vol.~22, pp.
  67--72, 2012.

\bibitem{Cordier1993}
P.. Cordier, M.. Mend\`es-France, P.. Bolon, and J.. Pailhous, ``Entropy,
  degrees of freedom, and free climbing: a thermodynamic study of a complex
  behavior based on trajectory analysis,'' \emph{International Journal of Sport
  Psychology}, vol.~24, pp. 370--378, 1993.

\bibitem{Cordier1994}
------, ``Entropy as a global variable of the learning process,'' \emph{Human
  Movement Science}, vol.~13, pp. 745--763, 1994.

\bibitem{Billat1995}
V.. Billat, P.. Palleja, T.. Charlaix, P.. Rizzardo, and N.. Janel, ``Energy
  specificity of rock climbing and aerobic capacity in competitive sport rock
  climbers,'' \emph{The Journal of Sports Medicine and Physical Fitness},
  vol.~35, pp. 20--24, 1995.

\bibitem{Seifert2013}
L.. Seifert, D.. Orth, R.. H\'erault, and K.. Davids, ``Affordances and
  grasping patterns variability during rock climbing,'' \emph{Studies in
  Perception and Action XII: Seventeenth International Conference on Perception
  and Action}, 2013.

\bibitem{Seifert2014}
L.. Seifert, L.. Wattebled, R.. H\'erault, G.. Poizat, D.. Ad\'e, N..
  Gal-Petitfaux, and K.. Davids, ``Neurobiological degeneracy and affordance
  perception support functional intra-individual variability of inter-limb
  coordination during ice climbing,'' \emph{PloS One}, vol.~9, 2014.

\bibitem{Fryer2012}
S.. Fryer, T.. Dickson, N.. Draper, M.. Eltom, L.. Stoner, and G.. Blackwell,
  ``The effect of technique and ability on the vo2–heart rate relationship in
  rock climbing,'' \emph{Sports Technology}, vol.~5, pp. 143--150, 2012.

\bibitem{White2010}
DJ.. White and PD.. Olsen, ``A time motion analysis of bouldering style
  competitive rock climbing,'' \emph{The Journal of Strength and Conditioning
  Research}, vol.~24, p. 1356, 2010.

\bibitem{Quaine1997}
F.. Quaine, L.. Martin, and JP. Blanchi, ``The effect of body position and
  number of supports on wall reaction forces in rock climbing,'' \emph{Journal
  of Applied Biomechanics}, vol.~13, pp. 14--23, 1997.

\bibitem{Pansiot2008}
J.. Pansiot, R.. King, D.. McIlwraith, B.. Lo, and G.. Yang, ``Climbsn: Climber
  performance monitoring with bsn,'' \emph{International Workshop on Wearable
  and Implantable Body Sensor Networks}, vol.~5, pp. 33--36, 2008.

\bibitem{Delignieres1993}
D.. Deligni\`eres, J.. Famose, C.. Th\'epeaut-Mathieu, and P.. Fleurance, ``A
  psychophysical study of difficulty rating in rock climbing,''
  \emph{International Journal of Sport Psychology}, vol.~24, pp. 404--416,
  1993.

\bibitem{Draper2011}
N.. Draper, T.. Dickson, G.. Blackwell, S.. Fryer, S.. Priestley, D.. Winter,
  and G.. Ellis, ``Self-reported ability assessment in rock climbing,''
  \emph{Journal of Sports Sciences}, vol.~29, pp. 851--859, 2011.

\bibitem{Madgwick2010}
S.. Madgwick, \emph{An efficient orientation filter for inertial and
  inertial/magnetic sensor arrays}, 2010.

\bibitem{Madgwick2011}
S.~O.~H.. Madgwick, A.~J.~L.. Harrison, and A.. Vaidyanathan, ``Estimation of
  imu and marg orientation using a gradient descent algorithm,''
  \emph{International Conference on Rehabilitation Robotics}, 2011.

\bibitem{Granjon2012}
P.. Granjon, \emph{The CUSUM algorithm, a small review}, 2012.

\bibitem{Minka}
T.P.. Minka, \emph{Estimating a Gamma distribution}, 2002.

\end{thebibliography}

\end{document}